\documentclass[aps,preprint,superscriptaddress,showpacs]{revtex4}

\usepackage{graphicx,epsfig}

\def\be{\begin{equation}}
\def\ee{\end{equation}}
\def\bea{\begin{eqnarray}}
\def\eea{\end{eqnarray}}

\def\case#1/#2{\textstyle\frac{#1}{#2}}
\def\k0{\kappa_{0}}

\begin{document}

\title{5D gravitational waves from complexified black rings}
\author{N. Bret\'on$^1$, A. Feinstein$^2$ and L. A. L\'opez$^{}$}
\affiliation{$^{}$ Dpto. de F\'{\i}sica, Centro de Investigaci\'on y de
Estudios Avanzados del I. P. N.,
Apdo. 14-740, D.F., M\'exico.\\
$^2$ Dpto. de F\'{\i}sica Te\'orica, Universidad del Pa\'{\i}s
Vasco, Apdo. 644, E-48080, Bilbao, Spain.}

\begin{abstract}
In this paper we construct and briefly study the 5D time-dependent solutions
of general relativity obtained  via double analytic continuation of the
black hole (Myers-Perry) and of the black ring solutions
with a double (Pomeransky-Senkov) and a single rotation (Emparan-Reall). The
new solutions take the form of a generalized Einstein-Rosen cosmology
representing
gravitational waves propagating in a closed universe.
In this context the  rotation parameters of the rings  can be interpreted as
the extra wave polarizations, while it is interesting to state that the
waves obtained from Myers-Perry Black holes exhibit an extra boost-rotational
symmetry in higher dimensions which signals their better behavior at null
infinity. The analogue to the C-energy is analyzed.
\end{abstract}

\pacs{04.30.-w, 04.50.Gh, 04.20.Jb, 11.10.Kk}

\maketitle

\section{Introduction}

The formulation of string theory in time-dependent backgrounds presents a
particularly challenging problem, although progress can be achieved by
considering some simple time-dependent solutions. As a step in this direction,
a class of time-dependent backgrounds has been investigated recently
\cite{Aharoni}; those spacetimes were obtained from a double analytic
continuation of asymptotically flat black holes, and describe the Lorentzian
evolution of a bubble. The technique of double analytic continuation was
originally developed to study the stability of the Kaluza–-Klein vacuum
\cite{Witten}, see also \cite{Dowker}. This technique has also been used in
the formulation of a positive energy theorem for anti-de Sitter space
\cite{Horowitz}, and discussed within the context of brane world scenarios
\cite{Ida}, and M-theory \cite{Costa-Fabi}. Similar techniques were also used
in \cite{Piran}, to obtain cylindrical gravitational waves with variable polarization
by starting with Kerr solution describing a rotating black hole.

The phenomena involving gravitational waves are under intense
investigation, in preparation for the upcoming expected observational data
provided by the gravitational wave detectors. Hystorically, the most often  
theoretically studied spacetime
interpreted as containing propagating gravitational radiation is that of the
Einstein-Rosen metric \cite{Carmeli}.

In this paper we are interested in studying a class of solutions obtained 
by a double analytic continuation of the known 5-D black ring spacetimes. 
Here, black ring
solutions are carried, by a Wick rotation, to generalized Einstein-Rosen
spacetimes. The obtained in such a way solutions  can be interpreted as cylindrical gravitational waves in 5 dimensions, as well as inhomogeneous 5-D universes. These may also serve as interaction regions of plane five dimensional gravitational waves which collide and focus on either a strong curvature singularity or a Cauchy horizon. The colliding wave re-interpretations are possible similarly to the standard 4-dimensional case due to the solitonic structure of the original solutions in the sense that the solitonic terms would provide the possibility of the continuation of the solutions into the plane wave regions \cite{FeinIb}. 
While the practical interest of the solutions in rather remote, they may serve as time-dependent backgrounds for superstring propagation, as well as can be easily connected to the so-called Pre-Big-Bang scenario solutions by dressing them with a massless dilaton field. They may also serve as test beds for numerical relativity.

A different issue we
address is that of the so called C-energy in radiative spacetimes; this
quantity, in certain sense, measures the amount of gravitational energy
carried by the waves towards infinity.

To transform the black ring solutions to the Einstein-Rosen form, we introduce
the technique of the double analytic continuation for metrics involving
polynomials of third and fourth degree as denominators;  in the process,
elliptic integrals and Jacobi functions come up.
 
In the following section II the double analytic continuation for the
Myers-Perry black hole solution is considered and  in Sec. III the solution
of black ring with a double rotation is continued analytically, obtaining an extra polarization for the 
gravitational waves. The subsequent section IV deals with the special case for which one of the rotation parameters is switched off  and the analogue of the Emparan-Reall ring is obtained. For all the cases we calculate the analogue to C-energy and present
plots  for the energy density and energy flux. Finally, some
conclusions are drawn in the last section.

\section{Gravitational waves from Myers-Perry black hole}

In this section, as a first step, the Myers-Perry 5D black hole solution \cite{Myers} is transformed via a Wick rotation into 
a hypercylindrical time-dependent spacetime.

We shall start with the Myers-Perry solution with one angular momentum,
in the coordinates $(\rho, \theta,t, \phi, \psi)$

\begin{equation}
ds^2=-dt^2+ \frac{\rho_0^2}{\Sigma} (dt-a \sin^2{\theta} d \phi)^2+ (\rho^2+a^2) 
\sin^2{\theta} d\phi^2 + \rho^2 \cos^2{\theta} d\psi^2 +
\frac{\Sigma}{\Delta} d\rho^2 + \Sigma d \theta^2,
\label{MP-metric}
\end{equation}

where $\Delta= \rho^2- \rho_0^2 + a^2$ and $\Sigma= \rho^2 + a^2 \cos^2{\theta}$.

With $\rho={\tilde R}+ \frac{\rho_0^2-a^2}{{ 4 \tilde R}}$, we transform 
the sector 

$\Sigma \left( \frac{d\rho^2}{\Delta} + d \theta^2 \right) \mapsto
\frac{\Sigma}{{\tilde R}^2} \left( d {\tilde R}^2 + {\tilde R}^2 d \theta^2 \right)$

Followed by the transformations

$r= {\tilde R} \sin {\theta}, \quad z= {\tilde R} \cos {\theta} ,\quad {\tilde R}^2=r^2+z^2$

the $({\tilde R}, \theta)$ sector is transformed into

$\frac{\Sigma}{{\tilde R}^2} \left( d {\tilde R}^2 + {\tilde R}^2 d \theta^2 \right) \mapsto  \frac{\Sigma}{{\tilde R}^2}(dr^2+dz^2).$

Applying now the Wick rotation $t \rightarrow i z$ and $z \rightarrow i t$, along with $a \mapsto ia$, the
metric (\ref{MP-metric}) becomes a time-dependent solution to the vacuum Einstein equations

\begin{eqnarray}
\label{MP-complex}
dS^{2}&=&\frac{\Sigma}{\tilde R^2}(dr^2-dt^2)+dz^2- \frac{\rho_0^2}{\Sigma}
\left( {dz-\frac{ar^2}{\tilde R ^2} d\phi} \right)^2
+ (\rho^2-a^2)\frac{r^2}{\tilde R ^2} d\phi^2 - \frac{\rho^2 t^2}{\tilde R ^2}d \psi^2 \nonumber\\
&=& \frac{\Sigma}{(r^2-t^2)}(dr^2-dt^2) + \gamma_{a b}dx^{a} dx^{b},
\end{eqnarray}
where $\Sigma= \rho^2 + \frac{a^2t^2}{{\tilde R}^2}$,
$\rho={\tilde R}+ \frac{\mu^2}{{ 4 \tilde R}}$, ${\tilde R}= \sqrt{r^2-t^2}$ and
$\mu^2 = \rho_0^2 +a^2$. Notice that $\psi$  now became a timelike coordinate that 
can be made spacelike with $\psi \to i \psi$. 
The line element
(\ref{MP-complex}) is of the form of a 5D generalized Einstein-Rosen metric,

\begin{equation}
ds^{2}=e^{f}(d\rho^{2}-dt^{2})+\gamma_{ab}dx^{a}dx^{b},
\end{equation}
where $x^{a}$ denote the Killing coordinates, $z, \phi, \psi$.

On the other hand, from \cite{harmark} we know that

det$\gamma_{ab}= - \frac{\rho^2}{4} \Delta \sin^2{2 \theta}=-\rho^2 \Delta \sin^2{\theta} \cos^2{\theta}$, which after the transformations becomes

\begin{equation}
{\rm det} \gamma_{ab}= \left( 1 - \frac{\mu^4}{16{\tilde R}^4} \right)^2 r^2t^2.
\end{equation} 
  
Being

\begin{equation}
G= \sqrt{{\rm det} \gamma_{ab}}= \left( 1 - \frac{\mu^4}{16{\tilde R}^4} \right)rt, 
\end{equation}
  
and the norm of the gradient of the transitivity surface
amounts to 
 
\begin{equation}
G_{\mu}G^{\mu}=\frac{r^2-t^2}{\Sigma}(G_{,r}^2-G_{,t}^2)= \frac{1}{\Sigma}
\left[ -(t^2-r^2)^2 -\frac{\mu^8-16^2 \mu^4 r^2t^2}{16^2 (r^2-t^2)^2}
+\frac{ \mu^4}{8} \right],
\label{grad1}
\end{equation} 

At this point it is convenient to perform yet another transformation  to null coordinates 
$(u,v)$ given by $u=r-t, v=r+t$, that puts the line element Eq. (\ref{MP-complex}))
into:

\begin{equation}
ds^2=  \frac{\Sigma}{uv}du dv + dz^2- \frac{\rho_0^2}{\Sigma} \left( 
dz+a \frac{(v+u)^2}{4uv} d\phi \right)^2 + (\rho^2 -a^2) \frac{(v+u)^2}{4uv}d \phi^2
- \rho^2 \frac{(v-u)^2}{4uv}d \psi^2,
\label{MPuv}
\end{equation}

where

\begin{equation}
\rho^2=uv+ \frac{\mu^2}{2} + \frac{\mu^4}{16uv}, \quad
\Sigma= \left( \sqrt{uv} + \frac{\mu^2}{4 \sqrt{uv}} \right)^2 + \frac{a^2(v-u)^2}{4uv}.
\end{equation}

Then in $(u,v)$-coordinates the norm $G_{\mu}G^{\mu}$ takes the form

\begin{equation}
G_{\mu}G^{\mu}= \frac{2uv}{\Sigma}G_{,u}G_{,v}= - \frac{1}{\Sigma}\frac{(\mu^4 -16u^4)(\mu^4 -16v^4)}{2(16)^2 u^2v^2}.
\label{grad2}
\end{equation} 
 
\begin{figure}
\centering
\includegraphics[width=8.6cm,height=6.5cm]{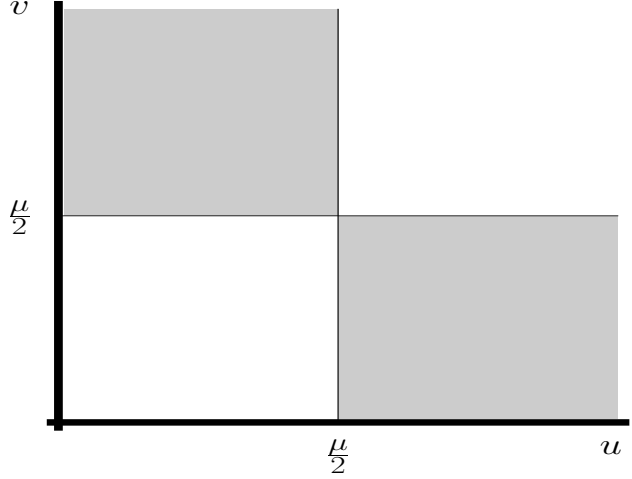}
\caption{\label{fig1}
The shaded regions correspond to the cylindrically symmetric
spacetimes, where $G^{\mu}G_{\mu} >0$. The regions are separated
by the lines $u=\mu /2$ and $v=\mu /2$ where $G^{\mu}G_{\mu} = 0$.
In the white regions $G^{\mu}G_{\mu} <0$}
\end{figure}

From the previous expression we see that the spacetime is clearly separated into four regions (see Fig. \ref{fig1}) 
where different interpretations
apply.

 
Since $\Sigma >0$, the sign of $G_{\mu}G^{\mu}$ depends on the ratios of
$u/\mu$ and $v/\mu$. These regions  are separated by
$u=\mu/2$ and $v=\mu/2$:  
For $u= \frac{\mu}{2}$ or $v= \frac{\mu}{2}$, $G_{\mu}G^{\mu}=0$
and it separates the regions where 
($u> \frac{\mu}{2}$ and $v > \frac{\mu}{2}$) 
or  ($u < \frac{\mu}{2}$ and $v < \frac{\mu}{2}$) where $G_{\mu}G^{\mu}<0$
and can be interpreted as gravitational waves propagating in a cosmological background; while
if ($u> \frac{\mu}{2}$ and $v < \frac{\mu}{2}$) 
or  ($u > \frac{\mu}{2}$ and $v < \frac{\mu}{2}$),
$G_{\mu}G^{\mu}>0$ that corresponds to cylindrical waves.

The previous statement can be posed in terms of trapped
surfaces.  Considering fixed  coordinates $x^{a}=\{ u,v\}$,
the invariant $\kappa$, \cite{Seno}, is given by

\begin{equation}
\kappa_{ u,v}= \frac{2 u^2v^2}{(u^2-v^2)^2} \frac{(\mu^4-16u^4)(\mu^4-16v^4)}{\Sigma (\mu^4-16u^2v^2)^2}
\end{equation}

We notice that the Killing orbits vanish at $u=v$ as well as at $u=-v$ corresponding to $t=0$ and $r=0$, therefore this spacetime has a singularity at $t=0$ as well as cylindrical axes at $r=0$. However, the singularity at $t=0$ is of a
Taub-NUT type, rather than a strong curvature singularity, and the spacetime may be extended across the hypersurface $t=0$ as in Taub-NUT, but the extension is not unique \cite{Misner}.
Moreover, whenever
$\kappa$ is everywhere negative (no zeroes),
the spacetime (\ref{MPuv}) does not possess neither marginally 
(no horizons), nor  trapped surfaces, 
therefore the solution (\ref{MPuv}) can be interpreted as cylindrical
gravitational waves in those regions. 
On the other hand, in the regions where $G_{\mu}G^{\mu}<0$ and
$\kappa >0$  trapped surfaces do exist with the possibility that  null congruences  converge to a singularity. Both regions are separated by the surfaces where
$u= \mu/2$ or $v= \mu/2$ .

In the 4D section of the regions where $G_{\mu}G^{\mu}>0$, therefore, we have a  cylindrically symmetric spacetime and the expression for the energy derived by
Brown-York \cite{Goncalves} should make sense.  One would also expect  the existence of a conical deficit along the angular coordinate.
Being the $E$-energy for the spacetime (\ref{MPuv})

\begin{equation}
\label{MP_E}
4E=1- e^{-f/2}= 1- \sqrt{ \frac{uv}{\Sigma}},
\end{equation}

At infinity $\Sigma \to \infty$, then as $v \to \infty$
(or $u \to \infty$), the energy
tends to a constant, $E(\infty)=1/4$. 
Accordingly, at infinity the energy flux and density,
$E_{,t}$ and $E_{,r}$, tend to zero.

Moreover, in the limit that $\mu \to 0$ ( $a \to 0$ and $ \rho_0 \to 0$)
the whole spacetime contains cylindrical gravitational waves, and  looks like

\begin{eqnarray}
ds^2&=&dudv+dz^2-\frac{(v-u)^2}{4}d \psi^2 + \frac{(v+u)^2}{4}d \phi^2 \nonumber\\
&=&dr^2-dt^2+dz^2 + t^2 d \psi^2 + r^2 d \phi^2,
\end{eqnarray}
Here we have complexified the $\psi$ coordinate to have a standard signature.
In fact, because of the above flat Kasner form of the 5-D metric, what one obtains here is the first example of the so-called 
boost-symmetric spacetime in higher dimensions \cite{Bicak}, see also \cite{Gowdy2}, and may represent radiation generated by accelerated sources in extra dimensions. It is interesting that if one generates cylindrical waves in a similar manner from a four dimensional black hole, no boost-symmetry is apparently present, it appears in the 5th-dimension due to the coupling of the radiation to the extra compact dimension-the term $t^2 d \psi^2$ in the line element in this limit.

\section{The double rotating black ring analytic continuation}

The starting point now is the black ring solution with two angular momenta  derived by
Pomeransky and Senkov (PS) \cite{pomeransky}. In coordinates $(x, y, t, \phi,
\psi)$ it is written as,

\bea
\label{PS_metric}
ds^2&=& \frac{H(y,x)}{H(x,y)}(dt+ \Omega)^2+\frac{F(x,y)}{H(y,x)} d\phi^2
+2 \frac{J(x,y)}{H(y,x)} d\phi d \psi \nonumber\\
&& - \frac{F(y,x)}{H(y,x)} d \psi^2 - \frac{2K^2 H(x,y)}{(x-y)^2(1- \nu)^2}
\left( \frac{dx^2}{G(x)} - \frac{dy^2}{G(y)} \right).
\eea
 
The solution represents the general  5D black ring solution with two
independent angular momenta (see analysis of PS solution in \cite{Elvang}).  
It depends on two coordinates: $-1 \le x \le 1$ and $- \infty < y < -1$. The
ranges of $t, \phi, \psi$ are: $- \infty < t < \infty$, $0 < \phi, \psi < 2
\pi$; the one-form $\Omega$ is given by

\bea
\Omega &=& - \frac{2K \lambda \sqrt{(1+ \nu)^2- \lambda^2}}{H(y,x)} \{
(1-x^2)y \sqrt{\nu} d \psi + \nonumber\\ && \frac{(1+y)}{(1- \lambda + \nu)}
[1+ \lambda - \nu +x^2 y \nu (1- \lambda - \nu)+2 \nu x (1-y)]d \phi \}
\nonumber\\ & = & \Omega_1 d \psi + \Omega_2 d \phi.  \label{Omega}
\eea

The functions $G, H, J, F$ are defined as:

\bea
G(x) &=& (1-x^2)(1+ \lambda x + \nu x^2), \nonumber\\ H(x,y)&=& 1+ \lambda^2 -
\nu^2 + 2 \lambda \nu (1-x^2) y + 2x \lambda (1 - y^2 \nu^2)+ x^2 y^2\nu (1-
\lambda^2 - \nu^2), \nonumber\\ J(x,y)&=& \frac{2K^2 (1-x^2)(1-y^2)\lambda
\sqrt{\nu}}{(x-y)(1- \nu)^2} [1+ \lambda^2 - \nu^2 + 2 \lambda \nu (x+y)  -x
y\nu (1- \lambda^2 - \nu^2)], \nonumber\\ F(x,y)&=& \frac{2K^2}{(x-y)^2(1-
\nu)^2} \{ G(x)(1-y^2)[((1-\nu)^2- \lambda^2)(1+\nu)+y \lambda (1- \lambda^2
+2 \nu-3 \nu^2)] \nonumber\\ &&+ G(y) [2 \lambda^2+ x \lambda((1-\nu)^2+
\lambda^2)+x^2((1-\nu)^2- \lambda^2)(1+\nu) \nonumber\\ && +x^3 \lambda (1-
\lambda^2 -3 \nu^2 +2 \nu^3) -x^4 (1-\nu) \nu (-1+ \lambda^2 + \nu^2)] \},
\label{metr_funcs}
\eea 

\begin{figure}
\centering
\includegraphics[width=8.6cm,height=6.5cm]{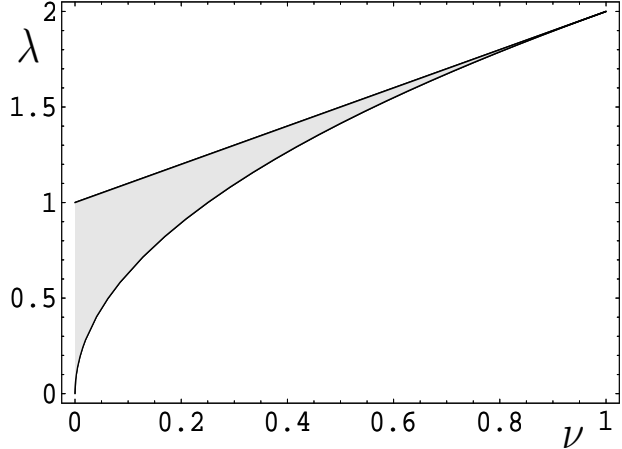}
\caption{\label{fig2}
Ranges for the parameters $\nu$ and $\lambda$ in the 
Pomeranski-Senkov solution, $0 \le \nu <1$ and $2 \sqrt{ \nu} \le \lambda < (1+ \nu)$,
the shaded region represents the $\lambda$ allowed values for a given $\nu$.}
\end{figure}

For a regular black ring, parameters $\lambda$ and $\nu$ must satisfy the
ranges $0 \le \nu <1$ and $2 \sqrt{\nu} \le \lambda < (1+ \nu)$, in order to
guarantee the existence of horizons, reality of the metric and positivity of
the black ring mass (see Fig. \ref{fig2}). The Emparan-Reall rotating black ring
\cite{emparan} is recovered when $\nu=0$. The interpretation of the solution
as a regular black ring is valid in the ranges where both, $G(x) >0$ ($-1 \le
x \le 1$) and $-G(y)>0$. The latter corresponds to two intervals:  $- \infty <
y < y_4$ and $y_3< y <-1$, where $y_3$ and $y_4$ are the two roots of $1+
\lambda y+ \nu y^2=0$.

 
To explore the time-dependent analogue  of the metric (\ref{PS_metric}), we
transform it into a 5D generalized Einstein-Rosen (ER) form, 

\be 
ds^2=
e^{f}(dR^2-dT^2) + \gamma_{ab}dx^{a}dx^{b}, \label{gER} 
\ee

where $\gamma_{ab}$ and $f$ are functions of $(R, T)$, and $x^{a}= z, \phi,
\psi$ are Killing directions. The local behaviour of the
solution is defined by the gradient of the element of the transitivity
surface, $G_{\mu}= \partial_{\mu} ({\rm det} \gamma_{ab})^{1/2}$ which can be
spacelike, timelike or null.  $G_{\mu} G^{\mu}$ is associated to the expansion
scalar and  defines the trapped  3-surfaces for fixed $(R, T)$. In case
$G_{\mu} G^{\mu} >0$, the solution can be interpreted as cylindrical
gravitational waves; when $G_{\mu} G^{\mu}$ varies from point to point then
(\ref{gER}) represents gravitational waves propagating along the $R$-direction
in an expanding universe.

In order to obtain the generalized ER form let us focus on the longitudinal 
sector to transform it as follows,

\be
\left( \frac{dx^2}{G(x)} + \frac{dy^2}{[-G(y)]} \right) \mapsto  (d R^2 + d \xi^2).
\label{nK_sector}
\ee
 
The next step would be to  perform the complex trick: $\xi \mapsto i T$
(this changes the sign of $d \xi^2$)  together with
$t \mapsto iz$ and $K \mapsto iK$; by doing so we 
arrive to the generalized Einstein-Rosen form (\ref{gER}).

Therefore, to begin with, we should perform the transformations,

\bea
\label{trans_x_y}
R &=& \int{\frac{d {\tilde x}}{\sqrt{G({\tilde x})}}}, \nonumber\\
\xi &=& \int{\frac{d{\tilde y}}{\sqrt{-G({\tilde y})}}}.
\eea 

The above integrals, where $G$ is a third or fourth-degree polynomial, are
known as the elliptic integrals of first, second or third kind of the
Legendre form \cite{handbook1}. The result of the integration depends 
on the roots of the polynomial inside the square root. The roots in our case
 are four real roots, $x_{1,2}= \pm 1$ and $x_{3,4}$ given by,
 
\be
x_{3,4}= - \frac{\lambda}{2 \nu} \pm \frac{\sqrt{\lambda^2-4 \nu}}{2 \nu}. 
\ee
Thus $G$ factorizes into:

\bea 
G(x)&=&(1-x^2)(1+ \lambda x + \nu x^2) \nonumber\\
&=& \nu (1-x)(x-(-1))(x-x_3)(x-x_4).
\eea 
  
Notice, that the roots may be ordered   $1> -1> x_3 > x_4$.
Moreover, since we are interested in a well defined coordinate transformation,
the integration (\ref{trans_x_y}) can not be performed over all the range of $
\tilde{x}$ or $\tilde{y}$, but rather only over intervals where the polynomials
$G(x)$ and $-G(y)$ are positive. These intervals are defined by two roots: $-1
< x <1$ for $x$ and $- \infty <y < y_4$ or $y_3 < y < -1$ for $y$.

For $x$ in the range $-1 \le x \le 1$ the apropriate integration turns out to
be \cite{handbook2}:
 
\bea 
R (x) &=& \int_{-1}^{x}{\frac{d{\tilde x}}{\sqrt{G({\tilde x})}}}=
\frac{2}{\sqrt{\nu (1-x_3)(-1-x_4)}} {\cal F}[\arcsin{\sqrt{\alpha}},
\sqrt{p}], \nonumber\\ \alpha&=& \frac{(1-x_3)(1+x)}{2(x-x_3)}, \quad p =
\frac{2 \sqrt{\lambda^2-4 \nu}}{\nu(1-x_3)(-1-x_4)}, \label{x_transf}
\eea
where ${\cal F}$ denotes the Legendre integral of first kind. 
The standard form to write this function is \cite{handbook2}
 
\be
z= \int^{\phi}{\frac{d \phi}{\sqrt{1-k^2 \sin^2{\phi}}}}={\cal F}[\phi,k],
\ee

where $\phi$ denotes the function amplitude of $z$, $\phi= {\rm am}(z)$
and the second argument of ${\cal F}$, $k$, is the moduli. 

With the transformation (\ref{x_transf}), the range of $R(x)$
is finite, $0< R(x)< R_{\rm max}$, $R_{\rm max}$ depends on the values 
of $\nu$ and $\lambda$, being $R_{\rm max}$ greater as $\nu$ increases.
The coordinate transformation for $y$ 
can be performed in the intervals where $-G(y) >0$,
through the integral

\be
\xi (y) = \int{\frac{d{\tilde y}}{\sqrt{-G({\tilde y})}}},
\ee
the appropriate intervals of integration are: $y_3 <y < -1$ or $ - \infty < y
< y_4$.  We note that as $\nu$ approaches zero, the interval $y_3 <y < -1$
becomes smaller. Therefore, let us choose the interval $ - \infty < y< y_4= -
\frac{\lambda}{2 \nu} - \frac{\sqrt{\lambda^2-4 \nu}}{2 \nu}$,

\bea 
\xi (y) &=& \int_{y}^{y_4}{\frac{d{\tilde y}}{\sqrt{-G({\tilde y})}}}=
\frac{2}{\sqrt{\nu(1-y_3)(-1-y_4)}} {\cal F}(\arcsin{\sqrt{\beta}}, \sqrt{q}),
\nonumber\\ \beta &=& \frac{(1-y_3)(y_4-y)}{(1-y_4)(y_3-y)}, \quad q =
\frac{\nu -1 + \sqrt{\lambda^2-4 \nu}}{\nu -1 - \sqrt{\lambda^2-4 \nu}}.  
\label{y_transf}
\eea

The range of $\xi$ is $ 0< \xi < \xi_{\rm max}$, where $\xi_{\rm max}$ 
depends on the values of $\nu$ and $\lambda$.

Having written in (\ref{x_transf}) and (\ref{y_transf}) the transformations $R(x)$,
$\xi(y)$, we should  now evaluate  the inverse transformations, $x(R)$ and $y(\xi)$ to
substitute into the functions depending on $(x,y)$ in order to express the metric
functions in terms of $(R, \xi)$.  The inverse functions involve the
Jacobi family of elliptic functions (sn$z$, cn$z$, dn$z$), with well
established analytical properties \cite{handbook1}, \cite{handbook2}. For
instance, the Jacobi function sn($z$) is defined by

\bea
\sin{\phi}&=& \sin({\rm am}z):= {\rm sn} (z), \nonumber\\
z&=& \int_{0}^{\phi}{\frac{d \phi}{\sqrt{1-k^2 \sin^2 \phi}}}={\cal F}[k, \phi],
\eea
and analoguosly are defined cn$(z)=\sqrt{1- {\rm sn}^2z}$ and
dn$(z)=\sqrt{1-k^2 {\rm sn}^2z}$ \cite{handbook1}. The Jacobi functions sn,
cn, dn, are real valued when their argument is real and the modulus $k$ is
either real or purely imaginary.

The specific inverse transformation depends again on the roots of the
polynomials $G(x)$ and $G(y)$ and are tabulated (see \cite{handbook2}
p.837). For the case we are dealing with they are

\bea
x(R)&=&\frac{-2 \nu-\lambda+\sqrt{\lambda^2-4 \nu}+2(\lambda -
\sqrt{\lambda^2-4 \nu}){\rm sn}^2(R,k') }{2 \nu+\lambda-\sqrt{\lambda^2-4
\nu}-(4 \nu) {\rm sn}^2(R,k')}, \ \label{X(R)}\\ y(\xi)&=&\frac{-2
\nu+\lambda+\sqrt{\lambda^2-4 \nu}+(2 \nu + \lambda + \sqrt{\lambda^2-4
\nu}){\rm sn}^2(\xi,k) }{-2 \nu+\lambda+\sqrt{\lambda^2-4 \nu}-(2
\nu+\lambda+\sqrt{\lambda^2-4 \nu} ){\rm sn}^2(\xi,k)}, \label{inverse_x_y}
\eea

where 

\be
k'^2=\frac{2 \sqrt{\lambda^2-4 \nu}}{- \nu+1+ \sqrt{\lambda^2-4 \nu}},
\quad
k^2=\frac{\nu-1+ \sqrt{\lambda^2-4 \nu}}{ \nu-1- \sqrt{\lambda^2-4 \nu}},
\ee

Going back to the non-Killing (longitudinal) sector (\ref{nK_sector}) that we have 
transformed from $(x,y)$ to $(R, \xi)$,

$\frac{dx^2}{G(x)} + \frac{dy^2}{-G(y)} \mapsto d R^2 + d \xi^2,$
 
we now proceed to perform the complex transformation $\xi \mapsto i T$.
Here we have to use the so-called  Jacobi's imaginary transformation,

\be
{\rm sn}(i \upsilon,k)=i \frac{{\rm sn}(\upsilon,k')}{{\rm cn}(\upsilon,k')},
\quad k'^2= 1-k^2.
\label{im_trans}
\ee
Notice that the expression for $y(\xi)$, Eq. (\ref{inverse_x_y}),
depends on ${\rm sn}^2$,
therefore, the transformation to imaginary argument shall give real functions,

\be
y(T)=\frac{(-2 \nu+\lambda+\sqrt{\lambda^2-4 \nu})({\rm cn}^2(T,k'))- (2 \nu +
\lambda + \sqrt{\lambda^2-4 \nu}){\rm sn}^2(T,k') }{(-2
\nu+\lambda+\sqrt{\lambda^2-4 \nu})({\rm cn}^2(T,k'))+(2
\nu+\lambda+\sqrt{\lambda^2-4 \nu} ){\rm sn}^2(T,k')},
\label{Y(xi)} 
\ee
 
Gathering all the results, as well as performing $t \mapsto iz$ and $K \mapsto
iK$, we write the double analytic continuation of the Pomeransky-Senkov
solution, in coordinates $(R, T, z, \phi, \psi)$ as

\bea 
\label{PS_complex}
ds^2&=& -\frac{H(y,x)}{H(x,y)}(dz+ \Omega)^2-\frac{F(x,y)}{H(y,x)} d\phi^2
-2 \frac{J(x,y)}{H(y,x)} d\phi d \psi \nonumber\\
&& + \frac{F(y,x)}{H(y,x)} d \psi^2 + \frac{2K^2 H(x,y)}{(x-y)^2(1- \nu)^2}
(d R^2 - d T^2),
\eea 
where $H$, $F$, $J$, $\Omega$ are given as in Eqs.
(\ref{Omega})-(\ref{metr_funcs})  and $x(R)$ and $y(T)$ are given by
expressions (\ref{X(R)}) and (\ref{Y(xi)}), respectively.

The volume of the transitivity surface, det$ \gamma_{ab}$, from Eq.
(\ref{PS_complex})  turns out to be

\be
{\rm det} \gamma_{a b}= \frac{F(x(R), y(T))F(y(T), x(R)) + J^2(x,y)}{H(x(R),
y(T))H(y(T), x(R))}.
\ee
 
Numerical results for the norm of the gradient of (det$\gamma_{a b}$),
indicate that it does not have the same sign throughout all the domain, but rather 
changes from point to point. Therefore, interpretation of the solution as
gravitational waves propagating on some cosmological  background is mandatory. We calculate  the Brown-York energy which is an analogue
of the C-energy,  that for the generalized ER
metric (\ref{gER}) is given by

\be
4E= 1- \exp{(- \frac{f}{2})},
\label{BY_en}
\ee
 
substituting from metric (\ref{PS_complex}) the Brown-York energy becomes

\be 
4E=1- \frac{(1- \nu)(x(R)-y(T))}{K \sqrt{2 H(x(R),y(T))}}.
\label{BY-energy}
\ee

This expression is not valid for all the range of $\nu$, since the
 metric function $H(x,y)$ vanishes in some places changing its sign throughout
the range of $(x,y)$. This behavior prevail after the coordinate
transformations; at those values of $x,y$ where $H(x,y)=0$, the energy
(\ref{BY-energy}) is ill-defined.  Also, one should take into account the
sign of $H(x,y)$ inside the square root in (\ref{BY-energy}). Numerically, we
obtain good behaviour for $\nu < 0.13$, although the ratio
$(x-y)/\sqrt{H(x,y)}$ behaves well for $\nu \le 0.2$. For values $\nu > 0.2$
the function $H(x,y)$ varies its sign frequently. The coordinate
transformations $x \mapsto R, \quad y \mapsto T$ introduce regions where
$x(R)=y(T)$, where the energy expression attains maxima. Generically the
energy diminishes as $\nu \to 1$ and grows as $K$ gets larger. In the limit
in which $\lambda \to (1+ \nu)$ the energy vanishes for large times. In the
limit that $\lambda \to 2 \sqrt{\nu}$ the Jacobi functions become the usual
trigonometric functions (for instance sn$(x) \to \sin(x)$) and the energy
oscillates throughout all the range of $(R,T)$.

\begin{figure}
\centering
\includegraphics[width=14cm,height=6.5cm]{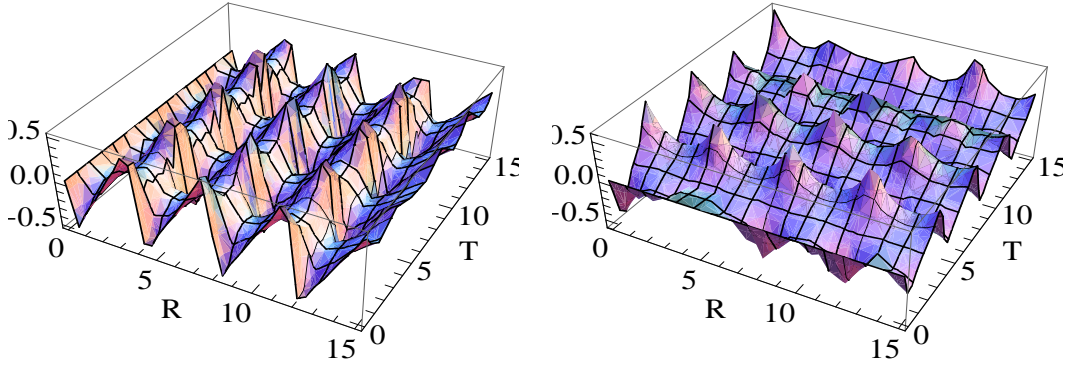}
\caption{\label{fig3}
Behaviour of the energy density (left) and
energy-flux (right), $\partial_{R}E, \partial_{T}E$, respectively,
for the values of $\nu =0.13, \quad K=1$ and $\lambda =0.75$
of the time-dependent analogous of Pomeranski-Senkov black ring}
\end{figure}

The plots (Fig. \ref{fig3}) show the local behaviour of the energy density and
energy-flux, $\partial_{R}E, \partial_{T}E$, respectively.  As $R \to
\infty, \quad T \to \infty$ the energy does not approach some definite limit but rather
oscillates, indicating that the gravitational waves have their origin in a singularity rather than in a localized source.
 
         
\section{Analytic continuation of the black ring with a single rotation}

The line element of the black ring of Emparan-Reall can be obtained
from (\ref{PS_metric}) if $\nu=0$. In this case the polynomials
$G(x)$ and $G(y)$ are of third order, and the roots are:

\be
G(x)=\lambda(1-x)(x-(-1))(x-(-1/  \lambda)).
\ee

Since $0< \lambda<1$, the root $x=-1 / \lambda$ is not included in the range
$-1 \le x \le 1$. Following the above outlined method, the transformation $x
\mapsto R$ is given by (\cite{handbook2}, p.219),

\bea
R &=& \int_{-1}^{x}{\frac{d{\tilde x}}{\sqrt{G({\tilde x})}}} \nonumber\\ &=&
\int_{-1}^{x}{\frac{d{\tilde x}}{\sqrt{\lambda (1-{\tilde x})({\tilde
x}+1)({\tilde x}+1/ \lambda)}}} \nonumber\\ &=& \frac{2}{\sqrt{\lambda +1}}
{\cal F}[\arcsin{\sqrt{\alpha}}, \sqrt{p}], \nonumber\\ \alpha&=&
\frac{(\lambda +1)(1+x)}{2(\lambda x+1)}, \quad p = \frac{2 \lambda}{\lambda
+1}. \label{x_transf3}
\eea
    
For the variable $y$, the root $y=1$ is not included in the range $-
\infty<y<-1$; then the interval for which $-G(y)>0$ is $-1/ \lambda \le y <-1$.
Thus the transformation $y \mapsto \xi$ is given by:

\bea
\xi &=& \int_{- \frac{1}{\lambda}}^{y}{\frac{d{\tilde y}}{\sqrt{-G({\tilde
y})}}} \nonumber\\ &=& \int_{- \frac{1}{\lambda}}^{y}{\frac{d{\tilde
y}}{\sqrt{\lambda (1-{\tilde y})(-1-{\tilde y})({\tilde y}+1/ \lambda)}}}
\nonumber\\ &=& \frac{2}{\sqrt{\lambda +1}} {\cal F}[\arcsin{\sqrt{\delta}},
\sqrt{m}], \nonumber\\ \delta&=& \frac{( y \lambda +1)}{1- \lambda}, \quad m =
\frac{1- \lambda}{1+\lambda}, \label{y_transf3}
\eea

Again ${\cal F}$ is the incomplete elliptic integral of the first kind. The
range of $\xi >0 $ is $0< \xi < 2 {\cal K} \sqrt{1- \lambda}/(1+ \lambda)$,
where ${\cal K}$ is the Legendre's complete integral of the first kind, ${\cal K}=
{\cal F}[\pi /2,k]$. Given $R(x)$ and $\xi(y)$, the inverse
transformations are, according to the intervals of integration (see
\cite{handbook2}, p. 837),

\bea
x(R)&=&\frac{-1 -\lambda + 2{} {\rm sn}^2(R,k) }{\lambda + 1-(2 \lambda) 
{\rm sn}^2(R,k)}, \label{x_ER}\\
y(\xi)&=& - \frac{1}{\lambda} +(\frac{1}{\lambda}-1){\rm sn}^2(\xi,k') \\
k^2&=& \frac{2 \lambda}{\lambda +1}, \quad
k'^2= \frac{1 - \lambda}{1 + \lambda}, 
\eea 

Performing now the complex transformation $\xi \mapsto i T$, and taking into account
(\ref{im_trans}), we obtain

\be
y(T)= - \frac{1}{\lambda} +(1-\frac{1}{\lambda})
\frac{{\rm sn}^2(T,k)}{{\rm cn}^2(T,k)}.
\label{y_ER}
\ee 
  
This is a real function. The ranges of the new variables are $ -\infty <R <
\infty$ and $T$ is in principle $ -\infty <T < \infty$ as well. However,
$y(T)$ diverges whenever cn$(T,k)=0$, that is for $T= \pm (2n-1){\cal K},
\quad n=1,2, \cdots$.  The result of the double analytic continuation,
$\xi \mapsto i T, \quad t \mapsto iz$ and $K \mapsto iK$, transforms
the line element (\ref{PS_metric}) with $\nu=0$ into

\bea
\label{ER_complex}
ds^2&=& - \frac{H(y)}{H(x)}(dz+ \omega d \phi)^2-\frac{F(x,y)}{H(y)} d\phi^2
\nonumber\\ && + \frac{F(y,x)}{H(y)} d \psi^2 + \frac{2K^2 H(x)}{(x-y)^2} (d
R^2 - dT^2),
\eea 
where $x(R)$ and $y(T)$ are given by expressions (\ref{x_ER}) and
(\ref{y_ER}), respectively. The metric function $J=0,$ while $\omega$, $G$,
$H$ and $F$ are given as (Eqs. (\ref{Omega})-(\ref{metr_funcs}) with $\nu=0$)

\bea 
\omega &=& - \frac{2K \lambda \sqrt{1- \lambda^2}}{H(y)}
\frac{(1+\lambda)}{(1- \lambda)} (1+ y), \nonumber\\ G(x) &=& (1-x^2)(1+
\lambda x), \nonumber\\ H(x)&=& 1+ \lambda^2 + 2x \lambda, \nonumber\\
F(x,y)&=& \frac{2K^2}{(x-y)^2} \{ G(x)(1-y^2)(1 + \lambda y)(1- \lambda^2)
\nonumber\\ &&+ G(y) [2 \lambda^2+ x \lambda(1+ \lambda^2)+ (x^2 +x^3
\lambda)(1- \lambda^2)] \},
\label{metr_funcs2}
\eea

The signature of spacetime is not aparent from the expression (\ref{ER_complex}).
We  have found out that the signs of the metric functions: $H(y(T)) < 0$, $H(x(R))>0$,
$F(x(R),y(T)) >0$, are independent of $\lambda$. On the other hand $F( y(T), x(R))>0$ if $
\lambda >0.5$ and for other values of $\lambda$ the expression for $F$ does not have a definite sign. The
term $[x(R)-y(T)]$ does not vanish anywhere  therefore after
the transformations, the metric remains regular.

The calculation of the element of the transitivity hypersurface gives in this case:

\bea 
{\rm det} \gamma_{ab}&=&g_{\psi \psi}(g_{zz}g_{\phi \phi}-g_{z \phi}^2) \nonumber\\
&=&\frac{F(x(R),y(T))F(y(T), x(R))}{H(y(T))H(x(R))},
\eea

\subsection{C-energy of the 5D polarized gravitational waves}

We  have evaluated numerically $\gamma_{\mu} \gamma^{\mu}$, the norm of the
gradient of the transitivity surface, $\gamma_{\mu}=\partial_{\mu} \sqrt{{\rm
det}\gamma_{ab}}$, with the result that it does not have a definite sign. Therefore, the
spacetime again, should be interpreted as gravitational waves propagating in a cosmology. The energy of
the system per unit Killing lenght $E$, can be calculated as the Brown-York
energy \cite{Goncalves}, Eq. (\ref{BY_en}), for the line element  (\ref{ER_complex}) gives

\be
4E=1 -  \frac{(x(R)-y(T))}{\sqrt{2K^2 H(x(R))}},
\ee

Since the functions $x(R)$ and $y(T)$ oscilate, $E$ has no 
clear limit at infinity, rather it oscillates all over the ranges of $R$ and
$T$. 
The energy density and energy flux are given, respectively, by

\be
E_{,R}=-\frac{1+\lambda^2+\lambda(x+y)}{4K(1+ \lambda ^2+2 \lambda
x)^{3/2}} x_{,R}, \quad
E_{,T}=- \frac{1}{4K \sqrt{2(1+ \lambda^2+2 \lambda x)}} y_{,T}.
\ee
 
\begin{figure}
\centering 
\includegraphics[width=14cm,height=6.5cm]{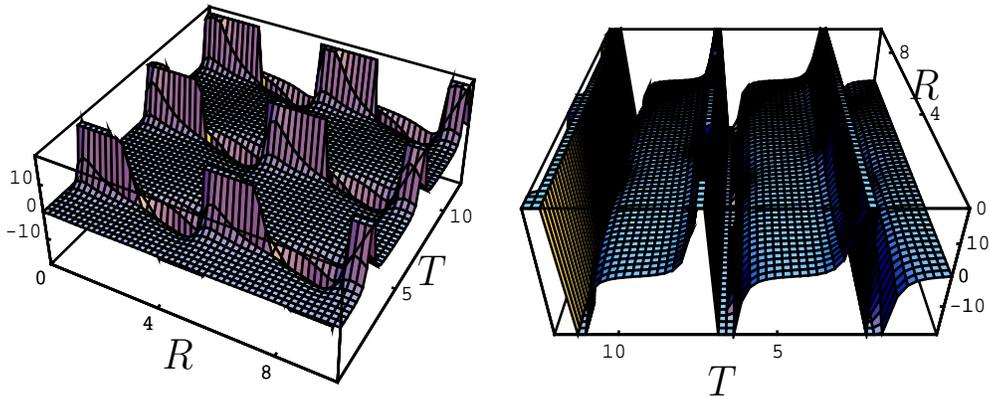}
\caption{\label{fig4}
Behaviour of the energy density (left) and
energy-flux (right), $\partial_{R}E, \partial_{T}E$, respectively,
for the time-dependent solution analogous to the black ring of Emparan-Reall.
The values of the parameters are $\lambda =0.5$, $K=1$}
\end{figure}

From the previous expressions (plots in Fig. \ref{fig4}) we can clearly observe the tendency that as $K$ (previously
interpreted as the rotation parameter of the ring and now stands for an extra polarization of the gravitational wave) gets larger, the energy does so too. Other
interesting limiting situation is at $R=0, T=0$, that corresponds to a finite energy,

\be
4E (R=0, T=0)= 1 -  \frac{1}{\sqrt{2}k \lambda}.
\ee

    
\section{Conclusions}

We note that the procedure of analytic continuation is not unique,
therefore, it may be that the presented solutions are not the only solutions one
is able to obtain by complexifying the black ring solutions.

The method presented to accomplish the complex analytic continuation works
well for solutions of this kind, i.e. having in  metric functions fourth
or third degree polynomials, leads to real solutions in a analytically continued metric. However,
interpretation turns out to be not a straight forward task, in that the coordinates
transform into linear combinations of Jacobi functions that are periodic. In
that sense the new coordinates $R,T$ should be compactified although the
ranges are initially taken to be $(- \infty, \infty)$. The situation here reminds closed inhomogeneous cosmological models studied in \cite{Carmeli2}.

Curiously enough, the analytic continuation of the Myers and Perry solution has a well defined hypercylindrical regions where the gradient of the transitivity element is always positive. This has to do with the boost-symmetric structure of the spacetime.
This does not happen in the other cases we have looked at, indicating that topologically these solutions have a complicated structure. It would be interesting to study the topology of these solutions in the future and to see as to whether one may impose special matching conditions as in Gowdy cosmologies \cite{Gowdy} on the hypersurfaces of vanishing transitivity element gradient in order to cure some pathologies. Also would be interesting to study further the differences between the analytic continuations of the black holes vs the continuations of the black rings. It looks as though the solutions obtained from black holes have a superior asymptotic structure:  due to their boost-rotational symmetry they inherit the future asymptotic structure of Minkowski space.
  
In the case of stationary generalizations of four-dimensional solutions one obtains a rich structure of singularities and event horizons. When these solutions are analytically continued into cylindrical and cosmological regions these structures are washed away. The $t=0$ singularities represent the inhomogeneous curvature singularity which is the source of the waves. The further evolution of the waves is conditioned by the overall metric expansion and is known to lead to a so-called Doroshkevich-Zeldovich-Novikov
DZN universe
\cite{Carmeli} filled with non-interacting null fluids. Incidentally, one may perform a dimensional reduction of these solutions along the extra coordinate to obtain an inhomogeneous universe with a massless scalar field. These have a well-understood singularity structure: either they start with a strong curvature singularity, or present a Cauchy horizon similar to the Taub-NUT universe. The $r=0$ quasiregular singularity, on the other hand signals the presence of an angular deficit and is related to the C-energy function. 

Finally, we  have detected that the Pomeransky- Senkov solution is not well defined throughout all
the original range $-1 \le x \le 1$ and $- \infty < y < -1$, in the sense that
metric functions (for instance, $H(x,y)$) changes sign and has zeroes besides
the  zeros considered as horizons; this makes problematic the
interpretation once the analytic continuation has been performed and restricts the range of the aplicability of the method.

\begin{acknowledgments} 
A.F.  acknowledges the support of the Basque Government Grant
GICO7/51-IT-221-07 and The Spanish Science Ministry Grant FIS2007-61800.
N. B. and L. A. L. thank the colleagues of UPV/EHU for warm hospitality.
L. A. L\'opez acknowledges Conacyt-M\'exico for a Ph. D. grant. Partial
support of Conacyt-Mexico Project 49182-F is also acknowledged.
\end{acknowledgments}

\end{document}